\documentclass[prb,showpacs,aps,twocolumn,nofootinbib]{revtex4}

\usepackage{graphicx,amssymb,amsmath}
\begin{document}

\title{Dephasing of Andreev pairs entering a charge density wave}       

\author{S. Duhot}
\affiliation{
Centre de Recherches sur les Tr\`es Basses
Temp\'eratures (CRTBT)\\ Bo\^{\i}te Postale 166, F-38042
Grenoble Cedex 9, France}
\author{R. M\'elin}
\affiliation{
Centre de Recherches sur les Tr\`es Basses
Temp\'eratures (CRTBT)\\ Bo\^{\i}te Postale 166, F-38042
Grenoble Cedex 9, France}

\begin{abstract}
An Andreev pair from a $s$-wave superconductor (S) entering a 
conventional gapless
charge density wave (CDW) below the Peierls gap dephases
on the Fermi wavelength while one particle states are localized
on the CDW coherence length. 
The paths following different sequences of impurities
interfere destructively, due to the different electron and hole densities
in the CDW. The same conclusion holds for averaging over
the conduction channels in the ballistic system.
We apply two microscopic approaches to this phenomenon:
i) 
a Blonder, Tinkham, Klapwijk (BTK) approach for 
a single highly transparent S-CDW interface; and
ii) the Hamiltonian approach for the Josephson effect in
a clean CDW and a CDW with non magnetic disorder.
The Josephson effect through a spin density wave (SDW) is
limited by the coherence length, not by the Fermi
wave-length.
\end{abstract}

\date{\today}
\maketitle

\section{Introduction} 
Condensed matter provides
many phases with an energy gap
between the ground state and the lowest excited state, and an
exponential decay of one-particle correlations.
Well-known examples of gapped (super)conducting or insulating phases
are superconductivity, the
quantum Hall effect, the Haldane gap in quasi-one dimensional
(quasi-1D)
spin $1$ antiferromagnets,
charge and spin density waves.
The coexistence between different orderings is usually
difficult in bulk systems, but the progress in
nanofabrication technology allows electron  
transport experiments in submicron
hybrids made of several electrodes with different
order parameters. 
Of particular interest are typical mesoscopic 
experiments with charge density waves (CDWs)
\cite{Bauer,Bauer-Josephson,Bobkova,Art-PRB}, 
such as transport through constrictions \cite{Neill},
through nanowires \cite{nanowire},
through an array of holes \cite{array},
through normal metal-charge density wave
(N-CDW) point contacts \cite{Kasatkin,Artemenko,AR},
an Aharonov-Bohm effect experiment \cite{AB},
and a scanning tunneling microscope experiment~\cite{Wang}.

Charge is transported below
the superconducting gap by Andreev reflection
at a normal metal-superconductor (N-S)
interface \cite{Andreev}:
a spin-up electron from the
normal side is reflected as a hole in the spin-down band and
a Cooper pair is transmitted in the superconductor \cite{Andreev}. 
Andreev pair transport through a 1D metallic channel was realized
recently in the form of the Josephson effect through
a carbon nanotube \cite{carbon}.

The tunneling current through an insulator
decays on the coherence length $\xi=\hbar v_F/\Delta$
(with $v_F$ the Fermi velocity and $\Delta$ the charge gap),
much larger than the Fermi
wavelength $\lambda_F$.
The dc-Josephson effect through a 1D channel
with translational symmetry breaking (a CDW) follows conventional
tunneling according to the first approach to this problem
by Visscher and Rejaei \cite{Bauer-Josephson}.
Coherent Andreev pair propagation can even be mediated
by the sliding motion of the CDW \cite{Bauer-Josephson,Yi},
suggesting that a mesoscopic CDW
can be depinned by a supercurrent.
On the other hand, Bobkova and Barash \cite{Bobkova}
found recently an absence of Andreev bound states at S-CDW-S interfaces.
We develop here 
a microscopic description of Andreev transport in S-CDW hybrids
based on the Hamiltonian approach,
successfully applied in the recent years to superconducting 
structures such as for instance a superconducting point
contact \cite{Cuevas}, ferromagnet-superconductor hybrids
\cite{Madrid}, and to non local transport through a superconductor
\cite{Deutscher,FFH,MF}.
Single particle evanescent states
are localized within the CDW coherence length $\xi_c=\hbar
v_F/|\Delta_c|$ (with $|\Delta_c|$ the Peierls gap),
much larger than $\lambda_F$. 
Andreev pairs are on the contrary
found to dephase on $\lambda_F$ in a CDW, compatible with
Ref.~\onlinecite{Bobkova} and
not captured by the quasiclassical
theory in Ref.~\onlinecite{Bauer-Josephson}.
This conclusion is also obtained from a Blonder, Thinkham,
Klapwijk (BTK) approach \cite{BTK,Bauer}.

The dephasing of a pair state
in a CDW is obtained in the same framework 
of the Hamiltonian approach as
non local Andreev reflection \cite{Deutscher,FFH,MF,Beckmann,Russo}
through a superconductor in N-S-N structures \cite{Russo},
a problem relevant to the realization \cite{Beckmann,Russo} of a source of
correlated pairs of electrons. The two problems are indeed dual:
the former is related to the propagation of an electron pair
through a CDW with electron-hole pairing, and the later is
related to the propagation in the electron-electron
and electron-hole channels through a superconductor with
electron-electron pairing.
The mechanism 
for non local transport through a superconductor consists however of
{\it opposite currents} in the electron-electron
and electron-hole channels because of the opposite sign of the charge carriers.
By contrast,
the effect in a CDW is an equilibrium property, that
we identify to the dephasing
 on $\lambda_F$ of the evanescent pair state.
The resulting absence of Josephson
effect through a CDW is robust,
independent of the interface
transparencies,
as opposed to being restricted to tunnel interfaces
in the superconducting case \cite{MF}.

The article is organized as follows.
A simple physical interpretation of the effect is presented in
Sec.~\ref{sec:phys}.
The microscopic model is presented in Sec.~\ref{sec:III}.
The BTK approach is presented in Sec.~\ref{sec:BTK}.
Boundary conditions at
interrupted chains and the supercurrent 
are discussed in Sec.~\ref{sec:Hamil}. 
Concluding remarks are given
in Sec.~\ref{sec:VI}.

\section{Physical picture}
\label{sec:phys}
Let us first consider non magnetic impurities in a CDW. 
For the sake of simplification,
the impurity potential is supposed to be weak enough for
the CDW phase to be the same as in the absence of impurities.
The discussion of localized phase deformations due to strong
pinning impurities is given in Sec.~\ref{sec:sol}.
Phase coherent Andreev reflection at normal metal-superconductor interfaces
implies that the backscattered hole in the normal metal follows
the same configure of impurities as the incoming electron, in
such a way as the different paths followed by 
an Andreev pair do not dephase with each other, except
for finite energy effects controlled by the Thouless energy,
and for inelastic scattering. 

By contrast, we show that in a CDW, the random phase factors
acquired by a spin-up electron visiting different impurities
do not cancel with the phase of a spin-down hole visiting the
same sequence of impurities, leading to dephasing of the Andreev
pair. The microscopic model discussed below in the ballistic system
shows that dephasing occurs on the smallest length scale:
the Fermi wave length, 
(up to a factor of two)
equal to the period of the CDW modulation.
The dephasing of an Andreev pair has its origin in the fact that
the total number of spin-up electrons at position $x$ along the chain,
given by
$N_{e,\uparrow}^{(CDW)}(x)=N_0+N_1\cos{(Qx+\varphi)}$
is dephased by $\pi$ compared to the
the total number of spin-down holes 
$N_{h,\downarrow}^{(CDW)}(x)=N_0'+N_1\cos{(Qx+\varphi+\pi)}=
N_0'-N_1\cos{(Qx+\varphi)}$, because a maximum in the
number of electrons corresponds to a minimum in the number of holes.
The quantities $N_0$, $N_0'$ and $N_1$ are respectively
the number of normal electrons and holes, and the amplitude of the
modulation in the number of electrons.
The total number of available states is $N_0+N_0'$.

Let us consider a single
non magnetic impurity in a CDW.
The impurity contribution to the energy of a spin-up electron
is
\begin{eqnarray}
\label{eq:HVe}
{\cal H}_{e,\uparrow}^{(imp)}&=&V(x_i) \left[N_{e,\uparrow}^{(CDW)}(x_i)
-N_0\right]\\
&=&V(x_i) N_1 \cos{(Q x_i+\varphi)}
\nonumber
\end{eqnarray}
equal to the same quantity for a spin-down hole:
\begin{eqnarray}
\label{eq:HVh}
{\cal H}_{h,\downarrow}^{(imp)}&=&-V(x_i)
\left[N_{h,\downarrow}^{(CDW)}(x_i)-N_0'\right]\\
&=&V(x_i) N_1 \cos{(Q x_i+\varphi)}
\nonumber
,
\end{eqnarray}
where $x_i$ is the 
position of the impurity and $V(x_i)$ the disorder impurity potential.
The $N_0$ and $N_0'$ terms that were subtracted the normal
ordered Eqs.~(\ref{eq:HVe})
and (\ref{eq:HVh})
induce an exactly opposite dephasing for an electron
and a hole following the same sequence of impurities, so these terms 
do not dephase the Andreev pair at equilibrium.
The $N_1$ terms are on the contrary
additive for electrons and holes making an Andreev pair
(see Eqs.~(\ref{eq:HVe}) and (\ref{eq:HVh})), resulting in
a dephasing between the different paths following different
sequences of impurities. 
As we show below by explicit calculations, the
same conclusion holds for a ballistic multichannel system,
where
the phase factors have their origin in Friedel oscillations. 

A spin density wave (SDW) can be described as two out-of-phase CDWs
for spin-up and spin-down electrons.
The number of spin-up electrons is $N_{e,\uparrow}^{(SDW)}(x)=N_0+
N_1 \cos{(Q x + \varphi)}$, and the number of spin-down electrons
is $N_{e,\downarrow}^{(SDW)}(x)=N_0-
N_1 \cos{(Q x + \varphi)}$. The total density is not
modulated, but the spin density is modulated.
The number of spin-down holes $N_{h,\downarrow}^{(SDW)}(x)=N_0+
N_1 \cos{(Q x + \varphi)}$ is equal to $N_{e,\uparrow}^{(SDW)}(x)$,
the number of spin-up electrons. We conclude
by the preceding argument that non magnetic
impurity random
phases of spin-up electrons and spin-down holes cancel with each
other in the total phase of the Andreev pair propagating through a
SDW, so that a 
Josephson effect over the coherence length is possible in a SDW.

We provide now three different microscopic
approaches to the absence of Andreev pair transport through a
ballistic CDW. Disorder is treated in Appendix~\ref{app:Born}.

\section{The model}
\label{sec:III}
The microscopic theory is based on the electronic part of the 1D
Peierls Hamiltonian of a ballistic CDW:
\begin{eqnarray}
\label{eq:H-tb}
{\cal H}&=&-\sum_{n,\sigma} \left(t_0+|\Delta_c| \cos{(2k_F x_n)}
\right) \times\\
&& \left( c_{n+1,\sigma}^+ c_{n,\sigma} +
c_{n,\sigma}^+ c_{n+1,\sigma}\right)
-\mu \sum_{n,\sigma} c_{n,\sigma}^+ c_{n,\sigma}
,
\nonumber
\end{eqnarray}
where $t_0$ is the mean hopping amplitude, 
$k_F$ the Fermi wave-vector, and
$\mu$ the chemical potential.
The summation over the integer $n$ runs over the sites
of the 1D chain. We have $x_n=n a_0$, with $a_0$ the
lattice parameter in the absence of the CDW modulation.
We suppose an incommensurate charge
density wave, unless specified otherwise in
the discussion of edge states.

\section{Blonder, Tinkham, Klapwijk approach}
\label{sec:BTK}
We evaluate now within the BTK
approach \cite{BTK}
subgap transport at a S-CDW interface, which was already probed
experimentally in Ref. \cite{Sin}.
A BTK approach to N-CDW interfaces
can be found in Ref.~\onlinecite{Bauer}. A scattering approach
to S-CDW interfaces with unconventional superconductors and
charge density waves can be found in Ref.~\onlinecite{Bobkova}.
To describe
the CDW and superconducting correlations on an equal footing,
we introduce a four component wave-function
corresponding to the four creation and annihilation operators
$c_{k,R,\uparrow}^+$, $c_{k,L,\uparrow}^+$,
$c_{k,R,\downarrow}$ and $c_{k,L,\downarrow}$
of right (R) and left (L) moving
spin-$\sigma$ electrons ($\sigma=\uparrow,\downarrow$)
of wave-vector $k$.
The wave-function
in the CDW part of the junction is given by
\begin{eqnarray}
&&\psi_{CDW}(x<0)=\left(\begin{array}{c}u e^{i\varphi}
\\v\\0\\0\end{array}\right) e^{(ik_F +{1\over\xi_c})x} \\
\nonumber
&+&
b \left(\begin{array}{c}ue^{i\varphi}
\\v\\0\\0\end{array}\right) e^{-(ik_F -{1\over\xi_c})x} 
+
b'
\left(\begin{array}{c}v^*\\u^* e^{-i\varphi}
\\0\\0\end{array}\right) e^{(ik_F +{1\over\xi_c})x}\\
\nonumber
&+&
a
\left(\begin{array}{c}0\\0\\ue^{i\varphi}
\\v\end{array}\right) e^{(ik_F +{1\over\xi_c})x}
+ a'
\left(\begin{array}{c}0\\0\\v^*\\
u^*e^{-i\varphi}\end{array}\right)
e^{(-ik_F +{1\over\xi_c})x}
,
\end{eqnarray}
where $x$ is the coordinate along the chain,
$u$ and $v$ are the CDW coherence factors, and
$u^*$ and $v^*$ their complex conjugate.
The wave-functions in the superconductor take the form
\begin{eqnarray}
\nonumber
&&\psi_S(x>0)
=d\left(\begin{array}{c}u_0\\0\\v_0\\0\end{array}\right)
e^{(i k_F -{1\over\xi_s})x}\\
\nonumber
&+&
d' \left(\begin{array}{c}0\\u_0\\0\\v_0\end{array}\right)
e^{-(i k_F +{1\over\xi_s})x}+
c
\left(\begin{array}{c}v_0\\0\\u_0\\0\end{array}\right)
e^{-(i k_F +{1\over\xi_s})x}\\
\nonumber
&+&
c'
\left(\begin{array}{c}0\\v_0\\0\\u_0\end{array}\right)
e^{(i k_F -{1\over\xi_s})x}
,
\end{eqnarray}
where $u_0$ and $v_0$ are the BCS coherence factors.
Matching the wave-functions and their derivatives for highly
transparent interfaces leads to $c=d=c'=d'=a'=a=b=0$ and
$|b'|^2=1$. No charge is transported by the 
reflection of a right-moving CDW quasiparticle in a left-moving
quasiparticle, as for a N-CDW interface \cite{Kasatkin,Artemenko}.

A pair from the superconductor decomposes on
pairs of evanescent CDW quasiparticles. The resulting
forward and backward combinations are both spin singlets
but interfere destructively
in the CDW, in such a way as to produce an absence of Andreev
pair penetration in a CDW.

\begin{figure}
\begin{center}
\includegraphics [width=.8 \linewidth]{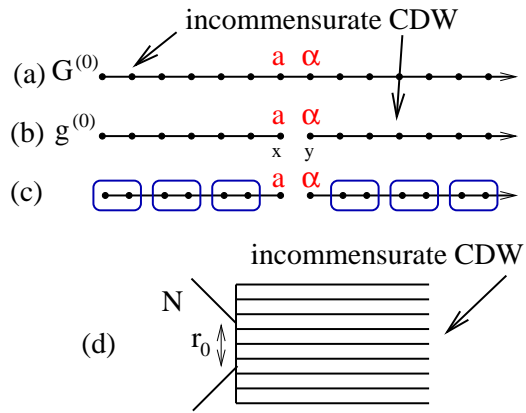}
\end{center}
\caption{Schematic representations of (a): a fully connected CDW chain;
(b): a disconnected chain; (c): a disconnected
dimerized state with zero energy edge state levels at sites
$a$ and $\alpha$ and a dimer order parameter along the chain \cite{Riera,FM};
(d): a junction of cross section
area $\sim r_0^2$ between a normal metal and an incommensurate CDW
with the chains perpendicular to the interface. 
See Ref.~\onlinecite{AR} for the experimental
realization of (d).
For clarity, (c) is drawn
for a dimerization, but the model applies to the more general case
of incommensurate modulations.
\label{fig:a}
}
\end{figure}

\section{Hamiltonian approach}
\label{sec:Hamil}
\subsection{Edge states}
\label{sec:IV}
We discuss now the boundary conditions at the extremity of
a finite chain before considering the supercurrent in Sec.~\ref{sec:V}.
A CDW of finite length is obtained by disconnecting
an infinite chain as in Figs.~\ref{fig:a}a and b \cite{Madrid}.
The Green's functions of an infinite CDW are evaluated in
Appendix~\ref{app:Green}.
The advanced Green's functions of the connected (disconnected) chain
are denoted by $G^{A,e,\uparrow}(\omega)$
$\left(g^{A,e,\uparrow}(\omega)\right)$
(see Fig.~\ref{fig:a}a).
They are related to each other by the Dyson equation
\begin{eqnarray}
G_{a,\alpha}^{A,e,\uparrow}(\omega)&=&
g_{a,\alpha}^{A,e,\uparrow}(\omega)\\
&+&
g_{a,a}^{A,e,\uparrow}(\omega)
t_{a,\alpha} G_{\alpha,\alpha}^{A,e,\uparrow}(\omega)
\nonumber
.
\end{eqnarray}
The condition that the chain
in Fig.~\ref{fig:a}b is disconnected is expressed
by $g_{a,\alpha}^{A,e,\uparrow}(\omega)=0$.
We use the notations
$t_{a,\alpha}= t_0+|\Delta_c| \cos{(2 k_F x)}$, 
$G_{a,\alpha}^{A,e,\uparrow}(\omega)=g^{A,e,\uparrow}_{x,y}
(\omega)$,
and $G_{\alpha,\alpha}^{A,e,\uparrow}(\omega)=
g^{A,e,\uparrow}_{y,y}(\omega)$,
$g^{A,e,\uparrow}_{x,y}(\omega)$
is defined by Eq.~(\ref{eq:g-tot-e-up}), 
and the neighboring sites $a$ and $\alpha$ are at coordinates
$x$ and $y$ (see Fig.~\ref{fig:a}a and b).
We deduce $g_{a,a}^{A,e,\uparrow}(\omega)=
G_{a,\alpha}^{A,e,\uparrow}(\omega)
/[t_{a,\alpha}
G_{\alpha,\alpha}^{A,e,\uparrow}(\omega)]$, leading to 
\begin{equation}
\label{eq:gaa}
g_{a,a}^{A,e,\uparrow}(\omega)\simeq \frac{\sin{(k_F a_0)} \sqrt{
|\Delta_c|^2-\omega^2}}{t_{a,\alpha}(\omega-\omega_0-i\eta)}
\end{equation}
for $\omega\simeq \omega_0$, where $\eta$ is small and positive.
We obtain a state of energy $\omega_0=
|\Delta_c| \cos{(\varphi_{x,y})}$, with
$\varphi_{x,y}=\varphi+k_F(x+y)$, localized
in a region of size $\xi_c$ at the extremity of the chain. 

We recover known results for a dimerized chain.
In the limit of a strong dimerization,
a semi-infinite chain ends either as a dimer or
as an isolated site (as in Fig.~\ref{fig:a}c),
resulting for the later in an edge state 
at zero energy 
at the extremity of each semi-infinite chain, corresponding
to $\varphi_{x,y}=\pi/2$ and $\omega_0=0$ in Eq.~(\ref{eq:gaa}).
This shows the consistency
between the Hamiltonian approach \cite{Madrid,Cuevas} 
used here for CDW hybrids, and the
known behavior of a dimerized system
\cite{Riera,FM}.

Considering now the incommensurate case,
the phase $\varphi_{x,y}$ entering the expression of
$\omega_0$ in Eq.~(\ref{eq:gaa})
is treated as a random variable
because of disorder in the position of the extremities of 
the chains at a multichannel N-CDW contact (see Fig.~\ref{fig:a}d).
A uniform
distribution of $\varphi_{x,y}$ leads to
a uniform subgap density of edge states at site $a$ (see Fig.~\ref{fig:a}):
$\rho_{a,a}(\omega)=\sin{(k_F a_0)}/\pi t_0$, as compared to
$\rho_{a,a}^N(\omega)=1/\pi t_0$ in the normal state.
The CDW pair amplitude at the extremity of the semi-infinite
chain is 
\begin{eqnarray}
\nonumber
&&F_{a,a}(|\Delta_c|,\omega)= \frac{|\Delta_c| \sin{(k_F a_0)}}
{t_{a,\alpha}^{(0)} \left[(\omega-i\eta)^2
- |\Delta_c|^2 \cos^2{(\varphi_{x,y})}\right]}\\
&\times&\left\{\omega \sin{(\varphi_{x,y})}
-\sqrt{ |\Delta_c|^2-\omega^2}
\cos{(\varphi_{x,y})} \right\}
.
\end{eqnarray}
The pair amplitude integrated over energy, and therefore the
self-consistent Peierls gap, vanish in a given
window of the phase $\varphi_{x,y}$,
leading to normal states 
at the CDW boundary because of interrupted chains.
The resulting subgap conductance at a N-CDW interface
in the geometry in Fig.~\ref{fig:a}d is
not contradicted by available experiments on N-CDW contacts \cite{AR}.
Normal states at the extremity of an interrupted
chain are also compatible with the appearance of a normal region
around a nano-hole induced by columnar defects
in a CDW film, which was proposed\cite{Bauer}
to explain
the Aharonov-Bohm oscillation experiment in a CDW \cite{AB}.

\begin{figure}
\begin{center}
\includegraphics [width=.8 \linewidth]{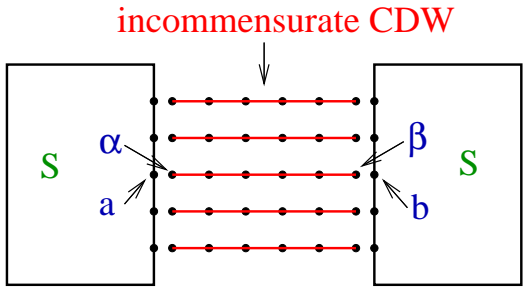}
\end{center}
\caption{Schematic representation of a S-CDW-S junction
between a superconductor (S), a charge density wave (CDW) and
another superconductor (S).
\label{fig:c}
}
\end{figure}

\subsection{Supercurrent}
\label{sec:V}
Now we evaluate the transport of Andreev pairs
through a CDW in the form of
the Josephson effect in
the device in Fig.~\ref{fig:c}.
The superconductors are described by
BCS theory, with a superconducting gap much smaller than the CDW gap, as
in possible experiments. 
The dc-supercurrent per conduction channel
for arbitrary interface transparencies is given by
\begin{equation}
\label{eq:super-exact}
I_S=\frac{e}{2\hbar} \int_0^{+\infty}
\frac{d\omega}{2i\pi}
\mbox{Im} \left\{
\mbox{Tr} \left\{ \hat{\sigma}_3
\left[\hat{t}_{a,\alpha} \hat{G}^A_{\alpha,\alpha}
,
\hat{t}_{\alpha,a} \hat{g}^A_{a,a} \right] \right\}
\right\}
\end{equation}
where $\left[\,,\right]$ is a commutator, 
$\hat{\sigma}_3$ is one of the Pauli matrices in Nambu space,
and the trace is a 
sum over the four components of the matrix Green's functions,
corresponding to spin degenerated right-left and electron-hole degrees of
freedom.

\subsubsection{Tunnel interfaces}
\label{sec:tunnel}
We first consider lateral atomic contacts connecting 
two superconductors to an infinite CDW. The
CDW lattice is supposed to be weakly modified by the contacts,
with no edge state.
The dc-supercurrent in the tunnel limit is
$I_S=I_c \sin{\varphi}$, with $\varphi$ the
superconducting phase difference, and 
with the critical current
\begin{equation}
\label{eq:super}
I_c=\frac{e}{h} \Delta_S T^2 t_0^2 \overline{
g^{A,e,\uparrow}_{\alpha,\beta}
(\Delta_S)
g^{A,h,\downarrow}_{\beta,\alpha}(\Delta_S)}
,
\end{equation}
where $T$ is the normal state transparency (with both the
CDW and the superconductor in the normal state),
$g^{A,e,\uparrow}_{\alpha,\beta}(\Delta_S)$
[see Eq.~(\ref{eq:g-tot-e-up})] and
$g^{A,h,\downarrow}_{\beta,\alpha}(\Delta_S)$ are the advanced
spin-up electron and spin-down hole
CDW Green's functions at energy $\omega=\Delta_S$, where
$\Delta_S$ is the superconducting gap 
[$\alpha$ and $\beta$ are shown in Fig.~\ref{fig:c},
and $t_0$ is defined by Eq.~(\ref{eq:H-tb})].
Averaging over the Fermi oscillations
due to the large number of
conduction channels in parallel in the incommensurate case
is denoted by
an overline in Eq.~(\ref{eq:super}).
We deduce from
Eq.~(\ref{eq:g-tot-e-up}), and from a similar expression
for $g^{A,h,\downarrow}_{\beta,\alpha}(\Delta_S)$
that the Andreev pair propagator 
$
\overline{
g^{A,e,\uparrow}_{\alpha,\beta}
g^{A,h,\downarrow}_{\beta,\alpha}}$,
and hence the tunnel
supercurrent
vanish as soon as the distance between
the superconducting electrodes exceeds
$\lambda_F$ (see Appendix~\ref{app:Green}),
in agreement with the preceding sections.

\subsubsection{Arbitrary interface transparencies}
To describe arbitrary interface transparencies, we 
treat multiple scattering at each interface to all orders \cite{MF}
while neglecting multiple Andreev reflections \cite{Cuevas}
because of the damping of the propagation in the CDW.
Within this approximation,
the dressed $4\times 4$
Green's function $\hat{\cal G}_{\alpha,\beta}$ is
$
\hat{\cal G}_{\alpha,\beta}=\hat{\cal M}_\alpha^{-1} \hat{g}_{\alpha,\beta}
\hat{\cal N}_\beta^{-1}
$,
with
$\hat{\cal M}_\alpha = \hat{I}-\hat{g}_{\alpha,\alpha}
\hat{t}_{\alpha,a} \hat{g}_{a,a} \hat{t}_{a,\alpha}$ and
$\hat{\cal N}_\beta = \hat{I}-\hat{t}_{\beta,b} \hat{g}_{b,b}
\hat{t}_{b,\beta} \hat{g}_{\beta,\beta}$,
where $\hat{t}_{a,\alpha}$ [$\hat{t}_{b,\beta}$]
are the diagonal hopping amplitude matrices with elements
$t_{a,\alpha}$ ($t_{b,\beta})$
for electrons, and $-t_{a,\alpha}$  ($-t_{b,\beta}$)
for holes.
We deduce from Eq.~(\ref{eq:super})
that the supercurrent 
vanishes beyond $\lambda_F$ whatever the interface
transparencies.

\subsubsection{Sliding motion}
Let us suppose now that a finite voltage is applied between the
two superconductors connected by low transparency interfaces
to a sliding CDW \cite{Bauer-Josephson}. 
The ac-Josephson
effect is treated by a gauge transformation in which the time-dependent
superconducting and CDW phases $\varphi(t)$ and $\chi(t)$ at time $t$
are absorbed in the hopping 
matrix elements \cite{Cuevas,Bauer-Josephson}.
The supercurrent, obtained from the Keldysh Green's function,
is expanded as in the S-N-S case in terms of the harmonics
$G_{r,s}(\omega)=\tilde{G}(\omega+r\omega_0/2,\omega+s \omega_0/2)$
of the Fourier transform of the Green's function
$G(t,t')$,
with $\chi(t)-\varphi(t)=\omega_0 t$, where we supposed
$\chi(t)$ linear in $t$. 
The supercurrent contains
the gauge transformed
Andreev pair propagator in the CDW, limited by $\lambda_F$
as in the dc-case.

\subsubsection{Edge states}
A finite length in the CDW chain (see Fig.~\ref{fig:c}) also
does not restore longer range Andreev pair transport.
The Green's functions $g_{\alpha,\beta}$ and $g_{\beta,\alpha}$
of the finite chain
are indeed related to the Green's functions $G_{a,\beta}$
and $G_{\beta,a}$ of the infinite chain by
$g_{\alpha,\beta}=G_{a,\beta}/[G_{a,a} t_{a,\alpha}]$ and
$g_{\beta,\alpha}=G_{\beta,a}/[G_{a,a} t_{\alpha,a}]$
(see the notations in Fig.~\ref{fig:c}).
The supercurrent of the finite chain is proportional to
the Andreev pair propagator of the infinite chain, again limited by
$\lambda_F$. Normal states of range $\xi_c$ were obtained at
both interfaces if the CDW gap is self-consistent. The
pair current does however not propagate through the gapped region between 
the edges. This contrasts with the Josephson effect through 
a discrete state in a nanotube of finite length \cite{carbon}.

\subsubsection{Bulk defects}
\label{sec:sol}
A supercurrent may propagate through a perturbed region of
size $\xi_c$ around the defects \cite{Larkin,Ov,Gleisberg},
but does not propagate through the
regions with no defect.
Hopping between possible normal regions in the CDW \cite{Gleisberg}
does not restore a longer range Josephson effect.

\subsubsection{Weak disorder in the Born approximation}
Disorder treated in the Born approximation \cite{Artemenko-Volkov}
is relevant to the description of the effect of disorder 
on the sliding motion. As shown in Appendix~\ref{app:Born}, the
supercurrent vanishes also for disorder in the Born
approximation.

\section{Discussion and conclusion}
\label{sec:VI}
To conclude, we have shown that pair states
are dephased within $\lambda_F$ in a CDW,
in spite of one particle states localized on $\xi_c$.
The absence of Josephson effect through a CDW 
is an intrinsic property of a ballistic multichannel multichannel
or disordered single channel CDW,
robust against
highly transparent interfaces, finite voltages,
and disorder. The supercurrent through a single channel
CDW shows oscillations on a length scale $\sim \lambda_F$.
Hopping between
normal states due to a self-consistent CDW gap vanishing locally
because of edges or bulk defects,
does not allow to overcome the absence of pair propagation
in the gapped regions.
Transport of Andreev pairs is possible through a SDW.

Spin active interfaces induce a Josephson current through a
half-metal ferromagnet \cite{Eschrig,Klapwijk} because of the
propagation of superconducting correlations among pairs
of electrons with the same spin orientation \cite{Bergeret,revue,Sosnin}.
Spin active S-CDW interfaces, or magnetic scattering in
the CDW (see Ref.~\onlinecite{magnetic1,magnetic2}
for blue bronzes), would change the
spin-down Andreev reflected quasi-hole in a spin-up quasi-hole
while preserving charge,
therefore preserving the absence of Josephson current.

The absence of Andreev pair penetration in a charge density
wave that we discussed is compatible with the fact that,
in bulk systems,
superconductivity hardly coexists with charge density waves
on the same portion of the Fermi surface
for small disorder. Other mechanisms are likely to be
responsible for the experimentally observed 
coexistence between superconductors and CDWs
in a number of compounds \cite{Gabovich}.
For instance,
besides
a possible coexistence between superconductivity and
charge density wave in layered compounds
and similar structures \cite{Balseiro,Bilbro},
a bulk coexistence\cite{LV} is possible
in the presence of a sufficient non magnetic disorder \cite{Mostovoy}. 
Andreev pair transport through a CDW is also possible if
the concentration of
non magnetic impurities in the CDW is such that the
CDW gap becomes smaller than the superconducting gap
(see Appendix~\ref{app:Born}). 

A parallel can be drawn with ferromagnet-superconductor (FS) 
structures.
In mesoscopic structures,
the Josephson current through a
ferromagnet shows oscillations leading to
$\pi$-junctions \cite{Buzdin1}, and, in bulk
structures, the coexistence between superconductivity
and ferromagnetism can lead to a Fulde-Ferrel-Larkin-Ovchinnikov
(FFLO) state\cite{FF,LO} with a spatially modulated gap.
On the other hand, 
it was proposed that, in addition to a uniform superconducting gap,
a FFLO\cite{FF,LO}-like
superconducting gap modulated at the CDW wave-vector $Q$ could
coexist with the CDW. The spatial oscillations of the
Josephson current obtained in Sec.~\ref{sec:Hamil}
in mesoscopic S-CDW-S structures
can be viewed as the CDW counterpart of the oscillations of
the Josephson current in S-F-S structures \cite{Buzdin1} that
average to zero in the case of strong ferromagnets
because of their very short period.

Regarding the possibility of transporting Andreev pairs
through a spin density wave that we obtained here, 
the compatibility between $s$-wave superconductivity and
SDWs can be seen in bulk systems from
the transition between a SDW and a
superconductor in the phase diagram of
the series of compounds (TMTTF)$_2$X and
(TMTTSF)$_2$X under pressure \cite{Jerome}.

From the point of view of experiments, the model of edge
states discussed above is a possibility for interpreting
the conductance spectra of N-CDW point contacts \cite{AR}. 
The realization of a point contact between a superconductor and a 
CDW is possible with the same technology as
in Ref.~\onlinecite{AR} for a N-CDW 
point contact, but with lower temperatures.
We expect a normal current as in Ref.~\onlinecite{AR}
if the temperature is such that
the superconductor is in the normal state
or if the applied voltage is larger than the superconducting
gap.
From our model,
no coherent transport of Andreev pairs is possible
in the CDW if the voltage is within the superconducting gap.
This is compatible with the sharp
increase of differential resistance within the superconducting
gap observed experimentally
in Fig. 1 c,d,e of Ref. \onlinecite{Sin}
for highly transparent Nb-NbSe$_3$ point contacts.
The realization of Josephson junctions
with CDWs or SDWs,
more technically involved, requires a very short distance
between the superconducting electrodes.
Finally, we note that, interestingly, two interacting particles
on a one dimensional quasi-periodic lattice 
lead to two-particle localized states
with quasi-delocalized one particle states \cite{Bellissard}.

\section*{Acknowledgments}
The authors acknowledge fruitful discussions with
J.P. Brison,
J. Dumas,
M. Houzet,
Y. Latyshev, J.C. Lasjaunias, P. Monceau , P. Rodi\`ere
and A.A. Sinchenko.
The Centre de Recherches sur les
Tr\`es Basses Temp\'eratures is associated with
the Universit\'e Joseph Fourier.

\appendix

\section{Green's functions of a charge density wave}
\label{app:Green}
The elements of the  $2\times 2$
advanced Green's function \cite{Artemenko-Volkov}
of a spin-up electron,
connecting two lattice
sites at positions $x$ and $y$ at energy $\omega$
with respect to the chemical potential $\mu$
are denoted by $g^{A,e,\uparrow,(i,j)}_{x,y}(\omega)$,
with $i,j= R,L$, corresponding 
to right and left moving fermions respectively.
Evaluating the
total advanced Green's function
obtained by summing over $i$ and $j$
leads to
\begin{eqnarray}
\label{eq:g-tot-e-up}
&&g^{A,e,\uparrow}_{x,y}(\omega)=\frac{1}{2 t_0} e^{-R/\xi_c(\omega)}
\left[ -\frac{\omega}{s(\omega)}
\cos{(k_F R)} \right.\\
&-&\left. \sin{(k_F R)}
+\frac{|\Delta_c|}{s(\omega)}
\cos{(\varphi+k_F(x+y))}\right]
\nonumber
,
\end{eqnarray}
with $R=x-y$,
$\Delta_c=|\Delta_c| \exp{(i\varphi)}$ the
complex CDW order parameter,
and $s(\omega)=\sqrt{|\Delta_c|^2-\omega^2}$.
The absence of Josephson effect discussed in Sec.~\ref{sec:tunnel}
for tunnel interfaces is obtained by noting that
the Green's function of a spin-down hole is
\begin{eqnarray}
\label{eq:g-tot-h-do}
&&g^{A,h,\downarrow}_{x,y}(\omega)=\frac{1}{2 t_0} e^{-R/\xi_c(\omega)}
\left[ -\frac{\omega}{s(\omega)}
\cos{(k_F R)} \right.\\
&-&\left. \sin{(k_F R)}
-\frac{|\Delta_c|}{s(\omega)}
\cos{(\varphi+k_F(x+y))}\right]
\nonumber
,
\end{eqnarray}
where the hole CDW phase been has changed \cite{note-signe}
by $\pi$ compared to the electron CDW phase in Eq.~(\ref{eq:g-tot-e-up})
(see Sec.~\ref{sec:phys}).

Averaging over the conduction channels leads to
\begin{equation}
\overline{g^{A,e,\uparrow}_{x,y}(\omega)
g^{A,h,\downarrow}_{y,x}(\omega)}=0
\end{equation}
for $|x-y|$ exceeding $\lambda_F$.

By contrast, in the SDW case, the Andreev pair propagator reduces to
$\overline{\left[g^{A,e,\uparrow}_{x,y}(\omega)\right]^2}$ because
the phase of spin-down electrons is shifted by $\pi$ compared to
the CDW case.
We find easily
\begin{equation}
\overline{\left[g^{A,e,\uparrow}_{x,y}(\omega)\right]^2}
=\frac{1}{4 t_0^2} \frac{|\Delta_c|^2}{|\Delta_c|^2-\omega^2}
\exp{\left(-\frac{2R}{\xi_s(\omega)}\right)}
,
\end{equation}
where $\xi_s(\omega)$ is the SDW coherence length. The
supercurrent through a SDW therefore decays over $\xi_s$,
not over $\lambda_F$ as for a CDW, in agreement with
Sec.~\ref{sec:phys}.

\section{Disorder in the Born approximation}
\label{app:Born}
The $2\times 2$ Green's functions
for the right-left components of a spin-up electron are given
by $\hat{G}=\hat{g}+\hat{g}\hat{\Sigma}\hat{G}$ in the Born approximation,
with the
self-energy
$\hat{\Sigma}=\int (dk/2\pi)
\hat{v} \hat{g}(k,\omega) \hat{v}^+$, where $\hat{g}(k,\omega)$
is the ballistic
matrix Green's function and $\hat{v}$ the matrix containing the
forward and backward scattering amplitudes.
Following Ref.~\onlinecite{Artemenko-Volkov}, we find
\begin{eqnarray}
\hat{\Sigma}&=& \int\frac{dk}{2\pi} \left[ |u|^2
\left(\begin{array}{cc}
g_{R,R}(k,\omega)& g_{R,L}(k,\omega) \\
g_{L,R}(k,\omega) & g_{L,L}(k,\omega)
\end{array} \right)\right.\\
&+&\left.
|v|^2
\left(\begin{array}{cc}
g_{L,L}(k,\omega)& 0 \\
0& g_{R,R}(k,\omega)
\end{array} \right) \right]
\nonumber
,
\end{eqnarray}
with $u$ and $v$ the amplitudes of backward and forward scattering.
We deduce the Green's function of a spin-up electron:
\begin{equation}
\label{eq:GA}
G^{A,e,\uparrow}(\xi_k,\omega)=
\frac{\overline{\omega}+\overline{\xi}_k\hat{\tau}_3
+\overline{\Delta_c}\hat{\tau}^+ + \overline{\Delta_c}^*
\hat{\tau}^-}
{\overline{\omega}^2-|\overline{\Delta_c}|^2-(\overline{\xi}_k)^2}
,
\end{equation}
with
\begin{eqnarray}
\overline{\Delta_c}&=&\Delta_c\left[1-\frac{1}{\tau s(\omega)}\right]\\
\overline{\omega}&=&\omega\left[1+\frac{1}{\tau s(\omega)}
+
\frac{1}{\tau' s(\omega)}\right]\\
\overline{\xi}_k&=&\xi_k+\alpha
,
\end{eqnarray}
with $\tau \overline{|u|^2}/\hbar v_F=1$, 
$\tau' \overline{|v|^2} /\hbar v_F=1$, 
$\xi_k$ the kinetic energy with respect to the chemical
potential, $\alpha$ a shift in the chemical potential,
and where
$\overline{\Delta_c}^*$ is the complex
conjugate of $\overline{\Delta_c}$. 
The matrices
$\hat{\tau}_3$, $\hat{\tau}^+$ and
$\hat{\tau}^-$ are given by
\begin{eqnarray}
\hat{\tau}_3=\left(\begin{array}{cc}
1 & 0 \\
0 & -1 \end{array} \right)\\
\hat{\tau}^+=\left(\begin{array}{cc}
0 & 1 \\
0 & 0 \end{array} \right)\\
\hat{\tau}^-=\left(\begin{array}{cc}
0 & 0 \\
1 & 0 \end{array} \right)
.
\end{eqnarray}
The Green's function of a spin-down hole is obtained by changing
$\Delta_c$ in $-\Delta_c$ as in Ref.~\onlinecite{Bauer-Josephson}.
An electron-hole transmission coefficient
is evaluated as in Ref.~\onlinecite{Smith}
for a superconductor. Eq.~(\ref{eq:GA}) leads to
$
\int (dk/2\pi) G^{A,e,\uparrow}(k,\omega)
G^{A,h,\downarrow}(k,\omega)=0
$,
obtained from evaluating the
matrix products and the integral over wave-vector.
This identity can be understood from the numerators of the
normal and anomalous contributions in
$G^{A,e,\uparrow}(\xi_k,\omega)G^{A,h,\downarrow}(\xi_k,\omega)$,
with the constraint 
$\overline{\omega}^2-|\Delta_c|^2-(\overline{\xi}_k)^2=0$.
We conclude that the Andreev propagator is limited by the
elastic mean free path since the transmission coefficient in the ladder
approximation vanishes after a single impurity scattering
coupling the spin-up electron to the spin-down hole
branches.


\begin{thebibliography}{99}
\bibitem{Bauer-Josephson} M.I. Visscher and B. Rejaei, Phys. Rev. Lett.
  {\bf 79}, 4461 (1997).

\bibitem{Bobkova} I.V. Bobkova and Yu.S. Barash,
  Phys. Rev. B {\bf 71}, 144510 (2005).

\bibitem{Bauer} M.I. Visscher and G.E.W. Bauer,
  Phys. Rev. B {\bf 54}, 2798 (1996);
  B. Rejaei and G.E.W. Bauer,
  Phys. Rev. B {\bf 54}, 8487 (1996);
  M.I. Visscher, B. Rejaei and G.E.W. Bauer,
  Phys. Rev. B {\bf 62}, 6873 (2000);
  
\bibitem{Art-PRB} S.N. Artemenko, Phys. Rev. B {\bf 67},
  125420 (2003).

\bibitem{Neill} K.O'Neill, E. Slot, R. Thorne and H. van der Zant,
J. Phys. IV France {\bf 131},
  221 (2005).
  
\bibitem{nanowire} E. Slot, M.A. Holst, H.S.J. van der Zant, and
S.V. Zaitsev-Zotov, Phys. Rev. Lett. {\bf 93},
  176602 (2004).
  
\bibitem{array} Yu. I. Latyshev, B. Pannetier
  and P. Monceau, Eur. Phys. J. B {\bf 3},
  421 (1998).
  
\bibitem{Kasatkin} A.L. Kasatkin and E.A. Pashitskii,
  Fiz. Nizk. Temp. {\bf 10}, 640 (1984);
  A.L. Kasatkin and E.A. Pashitskii,
  Fiz. Tverd. Tela (Leningrad) {\bf 27}, 2417 (1985)
  [Sov. Phys. Solid State {\bf 27}, 1448 (1985)].

\bibitem{Artemenko} S.N. Artemenko and S.V. Remizov, Pis'ma
  Zh. Eksp. Teor; Fiz. {\bf 65}, 50 (1997) 
  [JETP Lett. {\bf 65}, 53 (1997)].
  
\bibitem{AR} 
  A.A. Sinchenko, Yu. I. Latyshev, S. G. Zybtsev,
  I. G. Gorlova, P. Monceau, Pis'ma Zh. Eksp. Teor. Fiz. {\bf 64}, 259
  (1996) [JETP Lett. {\bf 64}, 285 (1996)];
  A.A. Sinchenko, Yu. I. Latyshev, S. G. Zybtsev, I. G. Gorlova,
  Zh. Eksp. Teor. Fiz. {\bf 113}, 1830 (1998)
  [JETP {\bf 86}, 1001 (1998)];
  Yu.I. Latyshev and A.A. Sinchenko, Pis'ma Zh. Eksp.
  Teor. Fiz. {\bf 75}, 714 (2002) [JETP Letters
    {\bf 75}, 593 (2002)].
  
\bibitem{AB} Yu. I. Latyshev, O. Laborde, P. Monceau, and
  S. Klaum\"unzer, Phys. Rev. Lett.
  {\bf 78}, 919 (1997).

\bibitem{Wang} Z.Z. Wang, J.C. Girard, C. Pasquier,
D. J\'erome, and K. Bechgaard, Phys. Rev. B {\bf 67}, 121401 (2003).

\bibitem{Andreev} A.F. Andreev, Zh. Eksp. Teor. Fiz {\bf 46},
  1823 (1964) [Sov. Phys. JETP {\bf 19}, 1228 (1964)];
  {\bf 49}, 655 (1966) [{\bf 22}, 455 (1966)].
  
\bibitem{carbon} P. Jarillo-Herrero, J.A. van Dam, and L.
Kouwenhoven, Nature {\bf 439}, 953 (2006).
  
\bibitem{Yi} J. Yi and S.-I. Lee, Phys. Rev. B {\bf 62},
9892 (2000).

\bibitem{Cuevas} J.C. Cuevas, A. Mart\'{\i}n-Rodero, and
  A. Levy Yeyati,
  Phys. Rev. B {\bf 54}, 7366 (1996).

\bibitem{Madrid} E. Vecino, A. Mart\'{\i}n-Rodero and
  A. Levy Yeyati, 
  Phys. Rev. B {\bf 68}, 035105 (2003).

\bibitem{Deutscher} G. Deutscher and D. Feinberg,
  Appl. Phys. Lett. {\bf 76}, 487 (2000).
  
\bibitem{FFH} G. Falci, D. Feinberg, and F.W.J. Hekking,
  Europhys. Lett. {\bf 54}, 255 (2001).

\bibitem{MF} R. M\'elin and D. Feinberg,
  Phys. Rev. B {\bf 70}, 174509 (2004);
  R. M\'elin, Phys. Rev. B
  {\bf 73}, 174512 (2006).

\bibitem{BTK} G.E. Blonder, M. Thinkham,
and T.M. Klapwijk, Phys. Rev. B {\bf 25}, 4515 (1982).

\bibitem{Beckmann} D. Beckmann, H.B. Weber, and H. v. L\"ohneysen,
  Phys. Rev. Lett. {\bf 93}, 197003 (2004).

\bibitem{Russo} S. Russo, M. Kroug, T.M. Klapwijk, and
  A.F. Morpurgo,
  Phys. Rev. Lett. {\bf 95}, 027002 (2005).

\bibitem{Riera} G.B. Martins, E. Dagotto, and
  J.A. Riera, Phys. Rev. B {\bf 54}, 16032 (1996).

\bibitem{FM} M. Fabrizio and R. M\'elin,
  Phys. Rev. Lett. {\bf 78}, 3382 (1997).

\bibitem{Larkin}  A.I. Larkin, Zh. Eksp. Teor. Fiz. {\bf 105}, 1793 (1994)
  [Sov. Phys. JETP {\bf 78}, 971 (1994)].
  
\bibitem{Ov} Yu. N. Ovchinnikov, K. Biljakovic,
  J. C. Lasjaunias and P. Monceau, Europhys. Lett.
  {\bf 34}, 645 (1996).

\bibitem{Gleisberg} S.N. Artemenko and F. Gleisberg,
  Phys. Rev. Lett. {\bf 75}, 497 (1995).

\bibitem{Artemenko-Volkov} S.N. Artemenko and A.F. Volkov,
  Zh. Eksp. Teor. Fiz. {\bf 80}, 2018 (1980)
  [Sov. Phys. JETP {\bf 53}, 1050 (1982)].

\bibitem{Eschrig} M. Eschrig, J. Kopu, J.C. Cuevas, G. 
Sch\"on, Phys. Rev. Lett. {\bf 90}, 137003 (2003).

\bibitem{Klapwijk} R.S. Keizer, S.T.B. Goennenwein, T.M. Klapwijk,
  G. Mia, G. Xiao, A. Gupta,
  Nature {\bf 439}, 825 (2006).

\bibitem{Bergeret} F. S. Bergeret, A. F. Volkov, and K. B. Efetov,
Phys. Rev. Lett. {\bf 86}, 4096 (2001).

\bibitem{revue} F. S. Bergeret, A. F. Volkov, and K. B. Efetov,
Rev. Mod. Phys. {\bf 77}, 1321 (2005).

\bibitem{Sosnin} I. Sosnin, H. Cho, V. T. Petrashov, and A. F. Volkov
Phys. Rev. Lett. {\bf 96}, 157002 (2006) .

\bibitem{magnetic1} J.Y. Veuillen, R. Chevalier, J. Marcus,
and C. Schlenker, Physica {\bf 143 B}, 186 (1986);
Solid State Comm. {\bf 63}, 587 (1987).

\bibitem{magnetic2} J. Dumas, B. Laayadi, and R. Buder,
Phys. Rev. B {\bf 40}, 2968 (1989).

\bibitem{Gabovich} A.M. Gabovich, A.I. Voitenko,
and M. Ausloos, Phys. Rep. {\bf 367}, 583 (2002).

\bibitem{Balseiro} C.A. Balseiro and L.M. Falicov,
Phys. Rev. B {\bf 20} (4457) (1979).

\bibitem{Bilbro} G. Bilbro and W.L. McMillan,
Phys. Rev. B {\bf 14}, 1887 (1976).

\bibitem{LV} P. B. Littlewood and C. M. Varma
Phys. Rev. Lett. {\bf 47}, 811 (1981) 
 
\bibitem{Mostovoy} M.V. Mostovoy, F.M. Marchetti, B.D. Simons,
and P.B. Littlewood, Phys. Rev. B {\bf 71}, 224502 (2005).

\bibitem{Buzdin1} A.I. Buzdin, L.N. Bulaevskii,
and S.V. Panyukov, Pis'ma Zh. Eksp. Teor.
Fiz. {\bf 35}, 147 (1982)
[JETP Lett. {\bf 35},
178 (1982)];
A. Buzdin, B. Bujicic, and M. Yu. Kupriyanov,
Zh. Eksp. Teor. Fiz.
{\bf 101}, 231 (1992)
[Sov. Phys. JETP {\bf 74}, 124 (1992)].

\bibitem{FF} P. Fulde and A. Ferrel,
Phys. Rev. {\bf 135}, A550 (1964).

\bibitem{LO} A. Larkin and Y. Ovchinnikov,
Zh. Eksp. Teor. Fiz. {\bf 47}, 1136 (1964)
[Sov. Phys. JETP {\bf 20}, 762 (1965)].

\bibitem{Jerome} P. Wzietek, F. Creuzet, C. Bourbonnais,
D. J\'erome, K. Bechgaard and P. Batail,
J. Phys. I France {\bf 3}, 171 (1993).

\bibitem{Sin} A.A. Sinchenko and P. Monceau,
J. Phys.: Condens. Matter {\bf 15}, 4153 (2003).

\bibitem{Bellissard} D.L. Shepelyansky, Phys. Rev. B
  {\bf 54}, 14896 (1996);
  A. Barelli, J. Bellissard, P. Jacquod, and
  D. Shepelyansky, 
  Phys. Rev. Lett. {\bf 77}, 4752 (1996).

\bibitem{note-signe} This is related to the
  change of sign of the Peierls gap between the electron
  and hole sectors (see Eq. (2) in
  Ref.~\onlinecite{Bauer-Josephson}).

\bibitem{Smith} R. A. Smith and V. Ambegaokar,
  Phys. Rev. B {\bf 45}, 2463 (1992).


\end{thebibliography}
\end{document}